\documentclass[twocolumn,epsfig,superscriptaddress,prl]{revtex4}

\usepackage{graphicx}
\usepackage{amssymb}
\usepackage{amsmath}
\usepackage{changes}
\usepackage{hyperref}
\usepackage{comment}

\begin{document}

\title{Majorana-induced DC Shapiro steps in topological Josephson junctions}	
\author{Sang-Jun Choi}
\email{sang-jun.choi@physik.uni-wuerzburg.de}
\affiliation{Institute for Theoretical Physics and Astrophysics, University of W\"urzburg, D-97074 W\"urzburg, Germany}
\author{Alessio Calzona}
\affiliation{Institute for Theoretical Physics and Astrophysics, University of W\"urzburg, D-97074 W\"urzburg, Germany}
\author{Bj\"orn Trauzettel}
\affiliation{Institute for Theoretical Physics and Astrophysics, University of W\"urzburg, D-97074 W\"urzburg, Germany}
\affiliation{W\"urzburg-Dresden Cluster of Excellence ct.qmat, Germany}
\date{\today}

\begin{abstract}
The demonstration of the non-Abelian properties of Majorana bound states (MBS) is a crucial step toward topological quantum computing. We theoretically investigate how Majorana fusion rules manifest themselves in the current-voltage characteristics of a topological Josephson junction. The junction is built on U-shaped quantum spin Hall edges and hosts a Majorana qubit formed by four MBS. Owing to Majorana fusion rules, inter- and intra-edge couplings among adjacent MBS provide two orthogonal components in the rotation axis of the Majorana qubit. We show that the interplay of the dynamics of the superconductor phase difference and the Majorana qubit governs the Josephson effect. Strikingly, we identify sequential jumps of the voltage across the junction with increasing DC current bias without external AC driving. Its role is replaced by the intrinsic Rabi oscillations of the Majorana qubit. This phenomenon, DC Shapiro steps, is a manifestation of the non-trivial fusion rules of MBS.
\end{abstract}
\maketitle

Majorana bound states (MBS) are non-Abelian excitations supported by topological insulators and represent the building blocks of topological quantum computation~\cite{Kitaev2001,Nayak2008,Alicea2011,Leijnse2012,Aasen2016}. The non-Abelian exchange statistic of MBS allows for the implementation of topological quantum gates processing quantum information in a topologically protected way~\cite{Ivanov2001,Bravyi2006}. A key property of MBS, which serves as an indirect demonstration of non-Abelian statistics, is their non-trivial fusion rules~\cite{Rowell2016,Nayak2008}. The latter implies that, under proper circumstances, the fusion of two MBS produces an equal-weight superposition of even and odd fermion parity, thus showing that MBS feature a non-trivial quantum dimension greater than one.

In condensed matter physics, several platforms capable of hosting and manipulating MBS have been investigated, including semiconducting quantum wires~\cite{Oreg2010, Lutchyn2010, Mourik2012, Albrecht2016, Deng2016, Guel2018} and more recently second-order topological superconductors~\cite{Yan2018, Hsu2018, Zhang2019, Zhang2020, Zhang2020b}. Among other platforms, topological Josephson junctions (TJJs) have proven to be promising theoretically and experimentally~\cite{Fu2008, Pientka2017, Choi2018, Guiducci2019, Fornieri2019, Calzona2019}. Importantly, single-electron tunneling into a pair of MBS within a TJJ leads to the fractional Josephson effect~\cite{Fu2009,Badiane2011,Jiang2011,San-Jose2012}, recently confirmed by experiments~\cite{Rokhinson2012, Wiedenmann2016, Bocquillon2017, Bocquillon2018, Laroche2019} with missing odd Shapiro steps. These results, together with other seminal experimental observations~\cite{Mourik2012,Albrecht2016,Guel2018,Wang2018}, have deeply strengthened the evidence for MBS. However, a compelling proof of their non-Abelian nature is still lacking.

\begin{figure}[b]
\centering
\includegraphics[width=0.99\columnwidth]{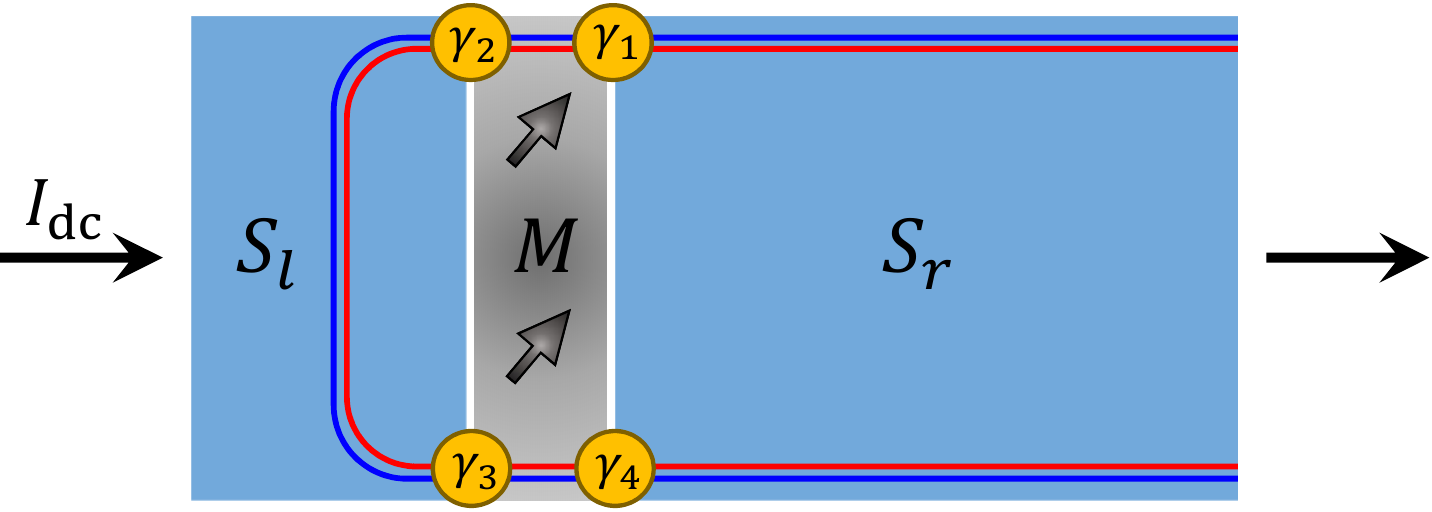}
\caption{ Josephson junction formed on U-shaped QSH edges hosting four MBS $\gamma_{1,2,3,4}$. Counter-propagating QSH edges are colored with blue and red lines.
The region with in-plane magnetization $\mathbf{M}$ (gray) constitutes weak links between two superconductors $S_l$ and $S_r$ (blue).
Pairs of two MBS at the upper and lower side of the junction compose two fermion parity states of $|0_{12}0_{34}\rangle$ and $|1_{12}1_{34}\rangle$, and a superposition of them defines a Majorana qubit state $|Q\rangle$. Intra-edge coupling provides the rotation of the Majorana qubit along the $z$-axis, and inter-edge coupling along the $x$-axis.
} \label{Fig:setupFig1}
\end{figure}

In this Letter, we present a novel Josephson effect which stems from the non-Abelian fusion rule of MBS. We consider a TJJ hosting four MBS defining a Majorana qubit (see Fig.~\ref{Fig:setupFig1}). Importantly, we allow a single Majorana bound state to fuse with different partners, for instance, $\gamma_2$ can develop an intra-edge coupling with $\gamma_1$ and/or an inter-edge coupling with $\gamma_3$. This leads to non-trivial dynamics of the Majorana qubit, whose interplay with dynamics of superconductor phase difference affects the Josephson effect. Remarkably, the Majorana dynamics lowers the critical current $I^*_c$ of the junction, and when a DC current bias increases over $I^*_c$, it induces sequential steps of the voltage drop $V$ across the junction.

We coin this phenomenon {\it{DC Shapiro steps}} to highlight the contrast to the known AC Shapiro steps. Both in conventional and topological Josephson junction, AC Shapiro steps appear (under certain conditions) if a DC current bias -- in combination with an AC current component -- is applied across the junction~\cite{Shapiro1963,Tinkham}. By contrast, the DC Shapiro steps featured by our setup emerge without any external periodic drive. The role of the AC driving is replaced by the intrinsic (Rabi) oscillations of the Majorana qubit, whose frequency can be estimated from the height of the steps. The importance of the novel DC Shapiro steps is two-fold: (i) They represent a new phenomenon in the realm of the Josephson effect. (ii) They are a manifestation of the non-trivial fusion rules of MBS.

{\it Setup---.} We consider the TJJ sketched in Fig.~\ref{Fig:setupFig1}, consisting of the edge of a quantum spin Hall (QSH) insulator which is proximitized by two superconductors separated by a ferromagnetic region. The Hamiltonian of the system is $H=\frac{1}{2}\int dx \Psi^\dagger(x) \mathcal{H}(x)\Psi(x)$ with
\begin{eqnarray}
\mathcal{H}(x) &=& \tau_z[-i\hbar v\partial_x s_z-\mu(x)] + \mathbf{M}(x)\cdot\mathbf{s} \nonumber\\
&& + \Delta_0(x)[\tau_x\cos\varphi+\tau_y\sin\varphi], \label{H}
\end{eqnarray}
where $\Psi^\dagger(x)=\left(\psi^\dagger_\uparrow(x),\,\psi^\dagger_\downarrow(x),\,\psi_\downarrow(x),\,-\psi_\uparrow(x)\right)$, and $\psi^\dagger_s(x)$ and $\psi_s(x)$ are creation and annihilation operators of electrons with spin index $s$. Pauli matrices $s_{x,y,z}$ and $\tau_{x,y,z}$ describe the spin and particle-hole space, respectively; $v$ is the Fermi velocity of the QSH edge states.
The chemical potentials in the region with magnetization and superconductors are, respectively, $\mu_M$ and $\mu_S$.
The proximity gap is induced only in the superconducting regions with $\Delta_0(x)=\Delta_0$. Regions $S_l$ and $S_r$ shall have a superconducting phase difference $\varphi$. We assume a finite magnetization $\mathbf{M}(x) = \text{sgn}(x)(M\cos\phi,M\sin\phi,0)$ only in the weak link between the superconductors. The distance between superconductors is $L$, and the length of the QSH-edge under the left superconductor is $W$. We consider that $\hbar v/W$ is much smaller than the bulk gap of the QSH insulator so that upper and lower edges are completely decoupled in the other regions. We focus on the Josephson effect at zero temperature.

To characterize the junction, we consider at first the limits  $\hbar v/L\ll M$ and $\hbar v/W\ll \Delta_0$, so that well-separated MBS $\gamma_{i=1,2,3,4}$ appear at zero energy. Those are equal superpositions of spin-polarized electrons and holes $\gamma_i\propto\int dx[\psi^\dagger_{i}(x)+\psi_{i}(x)]$ localized at position $x_i$, where
$\psi^\dagger_{i}(x) \propto e^{-|x-x_i|/\xi(x)}[\psi^\dagger_\uparrow(x) + e^{i\phi+i\theta_i(x)}\psi^\dagger_\downarrow(x)]$. The localization length
$\xi(x)$ is inversely proportional to the size of the gap of the region at position $x$. Importantly, the spin-polarization $\theta_i(x)$ is electrically tunable with chemical potentials: $\mu_M$ tunes $\theta_i$ at position $x_i$ according to $\sin\theta_i(x_i)=(-1)^i\text{sgn}(x_i)\sqrt{1-(\mu_M/M)^2}$;  $\mu_S$ changes the length of the spin helix of $\gamma_i$ in superconducting regions, i.e., $\theta_i(x) = \theta_i(x_i) + 2(x-x_i)\mu_S/(\hbar v)$. In the $M$-region, the spatial dependence of $\theta_i$ is $\theta_i(x)=\theta_i(x_i)$.

In order to enable the fusions of MBS, we now consider a different regime of length scales:
$\hbar v/L\sim M$ and $\hbar v/W\sim\Delta_0$. By projecting the full Hamiltonian $H$ on the Majorana wave-functions of $\gamma_i$, we obtain the effective low-energy Hamiltonian $H_\text{eff} = iE_x\gamma_2\gamma_3 + iE_z\gamma_1\gamma_2 + iE_z\gamma_3\gamma_4$. The inter-edge coupling through $S_l$ is $E_x \propto \sin\{[\theta_2(x)-\theta_3(x)]/2\}$, where $x$ is located in $S_l$. $E_x$ is vanishing (maximized) when the spin-polarization of $\gamma_2$ and $\gamma_3$ is (oppositely) aligned in the $S_l$-region. Explicitly, the inter-edge coupling is given by
\begin{eqnarray}
\label{eq:Ex}
E_x=\frac{2\Delta_0\sqrt{M^2-\mu_M^2}}{\Delta_0+\sqrt{M^2-\mu_M^2}}e^{-\frac{W\Delta_0}{\hbar v}} \sin\left(\frac{\mu_M}{M} + \frac{\mu_S W}{\hbar v}\right),
\end{eqnarray}
and the intra-edge coupling reads $E_z=E_M\cos\frac{\varphi}{2}$. The derivation of Eq.~\eqref{eq:Ex} is provided in the Supplemental Material~\cite{SM}. We stress that the presence of the magnetized region allows us to control the inter-edge coupling (via $\mu_M$) and to get rid of other mid-gap states at higher energies.

In order to study the Josephson effect, we describe the junction in terms of the complex fermions $f_{ij}\equiv(\gamma_i+i\gamma_j)/2$, which mediate the supercurrent. In particular, Majorana fermions $\gamma_1$ and $\gamma_2$ define two oppositely current-carrying states coined $|0_{12}\rangle$ and $|1_{12}\rangle$. They satisfy $f_{12}|0_{12}\rangle=0$ and $f_{12}^\dagger|0_{12}\rangle=|1_{12}\rangle$. Analogously, $\gamma_3$ and $\gamma_4$ provide current-carrying states $|0_{34}\rangle$ and $|1_{34}\rangle$. The total supercurrent across the junction is therefore controlled by the state of the Majorana qubit $|Q\rangle$. Importantly, states with total odd fermion parity do not carry any net supercurrent across the junction and they are invisible from a transport point of view~\cite{Trauzettel2014}. By contrast, the generic state with even total fermion parity $|Q\rangle=\alpha|0_{12}0_{34}\rangle+\beta|1_{12}1_{34}\rangle$ carries a finite supercurrent  $I_\text{J}= (|\alpha|^2-|\beta|^2)\frac{eE_M}{\hbar}\sin\frac{\varphi}{2}$. In the following, we focus on the even parity sector and recast $H_\text{eff}$ in terms of the current-carrying fermion states, $H_\text{eff} = E_x\sigma_x + E_z\sigma_z$, where the basis is $\{|1_{12}1_{34}\rangle\}, |0_{12}0_{34}\rangle\}$. We highlight that the off-diagonal elements, i.e., inter-edge coupling $E_x$, stem from the Majorana fusion rules $\sigma\times\sigma=I+\psi$~\cite{Nayak2008} and generate the Rabi oscillation between the current-carrying two-level states of the Majorana qubit.

{\it Josephson effect with Majorana dynamics---.} We consider an over-damped Josephson junction shunted by the normal resistor $R_N$. Under a DC current bias $I_\text{dc}$, the voltage drop $v(t)$ across the junction is described by the non-linear differential equation~\cite{Tinkham,Wiedenmann2016,Bocquillon2017}
\begin{equation}
I_\text{dc} = -\mathcal{Z}I_c\sin\frac{\varphi}{2} + \frac{\hbar}{2eR_N}\frac{d\varphi}{dt}, \label{Eq:Washboard}
\end{equation}
where $\mathcal{Z}=\langle Q|\sigma_z|Q\rangle$ and $I_c=eE_M/\hbar$ in our setup.
Dynamics of the phase $\varphi(t)$ is obtained by equating $I_\text{dc}$ with the sum of the supercurrent $I_\text{J}$ and the normal current $v(t)/R_N$, so that the voltage drop is determined by the Josephson relation $v(t)=\hbar/(2e)\dot{\varphi}(t)$. A mechanics analogue provides complementary insight to the dynamics of $\varphi(t)$: Eq.~(\ref{Eq:Washboard}) can be interpreted as the dynamics of a massless (phase) particle sliding along a tilted washboard potential $U(\varphi)=-I_\text{dc}\varphi +2\mathcal{Z}I_c\cos\varphi/2$~\cite{Tinkham}.

Without inter-edge coupling $E_x=0$, the Rabi oscillation between two current-carrying states is not produced, and $\mathcal{Z}$ is a conserved quantity: the profile of the associated washboard potential is fixed in time. In this case, a sudden change in $I_\text{dc}\le I_c$ causes $\varphi$ to evolve towards a fixed point $\varphi_{f}$ satisfying $I_\text{dc}=-\mathcal{Z}I_c\sin\varphi_f/2$. During the transient regime, $\varphi(t)$ monotonically increases with a characteristic time scale $\tau_{r}=\hbar/(2eI_cR_N)$. Thus, a positive voltage $v(t)$ is temporarily generated and vanishes afterwards. For a larger DC current, $I_{\rm dc}>I_c$, the phase $\varphi$ increases with time and a finite average voltage drop across the junction is developed $V=\lim_{\tau\rightarrow\infty}\frac{1}{\tau}\int_0^\tau v(t)dt$.

With a finite inter-edge coupling $E_x>0$, assumed to be positive without loss of generality, the behavior of the junction is controlled by the dynamics of both the phase particle and the Majorana qubit. To describe the latter, we introduce the unit vector $\mathbf{R}=(\mathcal{X}, \mathcal{Y}, \mathcal{Z})$ on the Bloch sphere, with $\mathcal{X}=\langle Q|\sigma_x|Q\rangle$ and $\mathcal{Y}=\langle Q|\sigma_y|Q\rangle$. Its dynamics is controlled by the equations of motion,
\begin{equation}
\frac{d\mathbf{R}}{dt} = \mathbf{N}\times\mathbf{R}, \label{Eq:MajoranaDynamics}
\end{equation}
where $\mathbf{N}=2(E_x,0,E_z)/\hbar$. If $E_{x/z}$ were time-independent, $\mathbf{R}$ would precess around the vector $\mathbf{N}$ with time period $\mathcal{T}=\pi\hbar/(E_x^2+E_z^2)^{1/2}$. However, as $E_z\propto\cos\varphi(t)/2$, the precession axis $\mathbf{N}(t)$ changes according to Eq.~\eqref{Eq:Washboard}, while the barrier height $\mathcal{Z}(t)$ in Eq.~\eqref{Eq:Washboard} changes simultaneously by Eq.~\eqref{Eq:MajoranaDynamics}.

\begin{figure}[b]
	\centering
	\includegraphics[width=0.99\columnwidth]{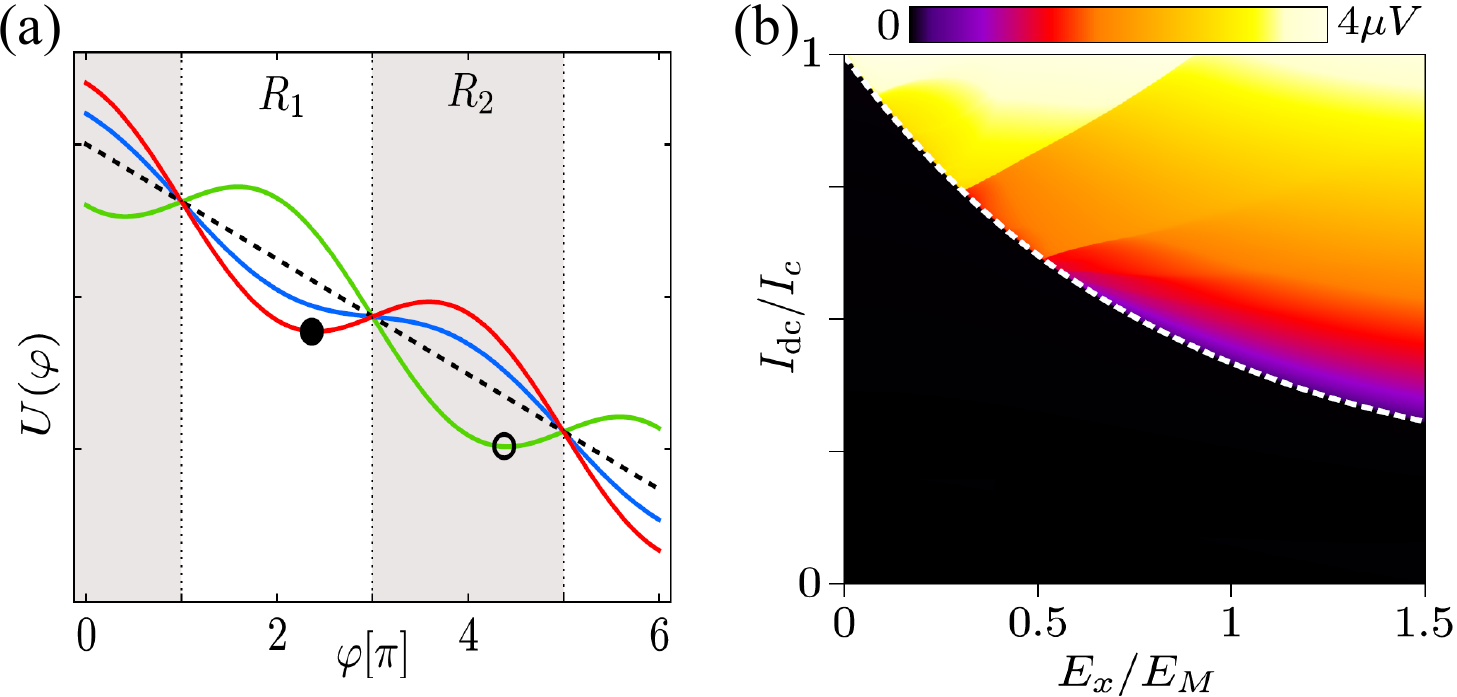}
	\caption{ (a) Sketch of the washboard potential $U(\varphi)$ when $\mathcal{Z}>\mathcal{Z}_\text{th}$ (red line), $0<\mathcal{Z}<\mathcal{Z}_\text{th}$ (blue line), $\mathcal{Z}=0$ (dashed line), and $\mathcal{Z}<-\mathcal{Z}_\text{th}$ (green line). When the Majorana qubit precesses with $\mathcal{Z}>0$, the phase particle (solid circle) is recaptured in $\varphi\in R_1=[\pi,3\pi]$. After recapturing the phase particle by a certain number of times $n$, $\mathcal{Z}$ flips its sign, and the phase particle rolls down to $\varphi\in R_2 = [3\pi,5\pi]$ (hollow circle).
	(b) Colormap of the time-averaged voltage $V$ under DC current bias $I_\text{dc}$ and the inter-edge coupling $E_x$. $I_c^*$ in Eq.~\eqref{Eq:Ic} is drawn as a guide for the eye (dotted line).  $V$ exhibits sequential jumps as increasing $I_\text{dc}$.
		 We use $E_M=5\,\mu\text{eV}$ and the experimental parameter of $2eI_cR_N/h\sim 2\,\text{GHz}$ in Ref.~\cite{Bocquillon2017}.} \label{Fig:Fig2dynamics}
\end{figure}

To develop intuition about the interplay between the differential equations~\eqref{Eq:Washboard} and~\eqref{Eq:MajoranaDynamics}, we focus on the features of the washboard potential $U(\varphi)$. As shown in Fig.~\ref{Fig:Fig2dynamics}(a) for a DC current $I_\text{dc}$, the qubit state $\mathcal{Z}$ modulates the profile of $U(\varphi)$ with the constant overall slope in time due to the absence of AC driving.
While $U(\varphi)$ is pinned at the points $\bar{\varphi}_j \equiv \pi\,(\text{mod}\, 2\pi)$, as for the regions $R_j: \bar{\varphi}_{j-1} \leq \varphi<\bar{\varphi}_{j} $ in between those points, two configurations are possible: (i) If $|\mathcal{Z}|$ exceeds the threshold $\mathcal{Z}_\text{th} = I_\text{dc}/I_c$, the regions feature an alternating pattern of local minima and maxima [red and green lines in Fig.~\ref{Fig:Fig2dynamics}(a)]. (ii) If $|\mathcal{Z}|<\mathcal{Z}_\text{th}$, the potential $U(\varphi)$ becomes a monotonically decreasing function [blue and dashed lines in Fig.~\ref{Fig:Fig2dynamics}(a)]. In configuration (i), the phase particle cannot leave the region of its initial location. Hence, it moves towards a local minimum. As this motion affects the direction of the precession axis $\mathbf{N}(t)$, the whole system starts to display damped oscillations because energy is dissipated via the resistance. If the bias current is sufficiently small $I_{\rm dc} \ll I_c$, the phase particle and the Majorana qubit eventually reach fixed points $\varphi_0$ and $\mathbf{R}_0$ as $t\rightarrow\infty$ and stop to move. Therefore, after a transient time, the average voltage drop vanishes $V=0$ and the ground state carries the supercurrent $I_{\rm dc}$ without resistance. 
We provide analytic solution of the fixed points and numerical results of $\varphi(t)$ and $\mathbf{R}(t)$ in Ref.~\cite{SM}.

As a physical consequence of finite inter-edge coupling $E_x$, we find that finite voltage develops as the bias current increases but still fulfills $I_{\rm dc}< I_c$. A finite value of $E_x$ stimulates the Majorana qubit to evolve from the state $|0_{12}0_{34}\rangle$ into a superposition of the opposite current-carrying states $|0_{12}0_{34}\rangle$ and $|1_{12}1_{34}\rangle$, corresponding to $|\mathcal{Z}|<1$. Hence, the configuration (ii) of the washboard potential $|\mathcal{Z}|<\mathcal{Z}_\text{th}=I_\text{dc}/I_c$ can occur and the phase particle can roll down. A sustained onset of this mechanism leads to a finite $V$ and therefore to a reduction of the critical current $I_c^*$ of the junction. The analytical computation of $I_c^*$ confirms the existence of this effect. In particular, by computing the maximal current carried by the ground state of the system, we obtain
	\begin{equation}
	\label{Eq:Ic}
	I_c^* = \sqrt{I_c^2+\left(\frac{e E_x}{\hbar}\right)^2}-\frac{e E_x}{\hbar} \leq I_c
	\end{equation}
which is indeed lowered by a finite inter-edge coupling $E_x$~\cite{SM}. In Fig.~\ref{Fig:Fig2dynamics}(b), we show the agreement between Eq.~\eqref{Eq:Ic} (white dashed line) and the numerical computation of $V$, which features finite values even for $I_{\rm dc}$ well below $I_c$. As the lowering of the critical current traces back to the fusion properties of the four MBS hosted by the TJJ, it represents the first key results of our work.

\begin{figure}[t]
	\centering
	\includegraphics[width = 0.99\columnwidth]{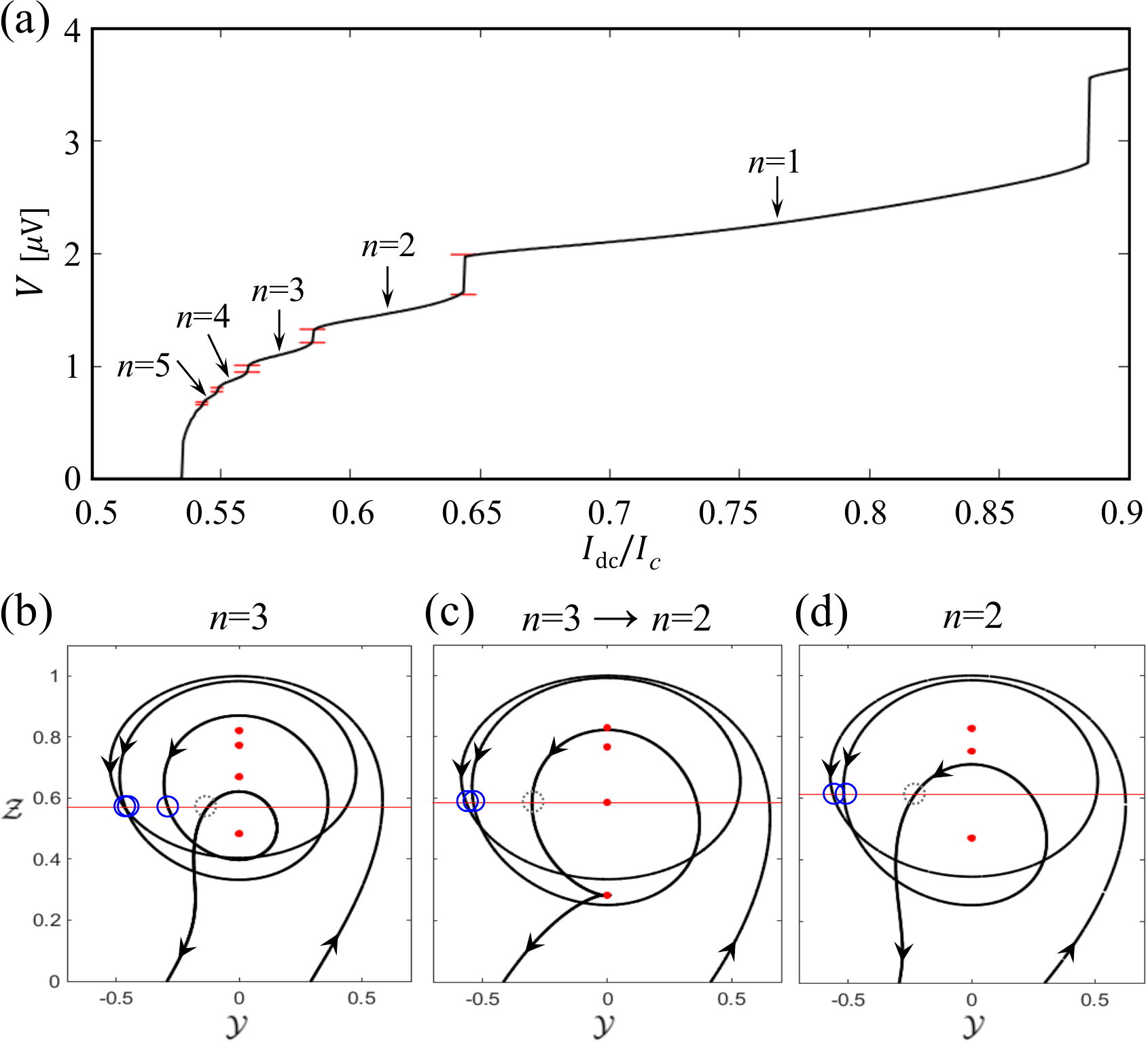}
	\caption{ (a) A current-voltage curve for $E_x/E_M=0.67$. The number $n$ of recapturing the phase particle is indicated and the analytically estimated size of jumps is shown with horizontal lines (red).
(b-d), the trajectories of $(\mathcal{Y}(t),\mathcal{Z}(t))$ are displayed for  various $n$. The Majorana qubit rotates around $\mathbf{N}$ with the counter-clockwise direction (black arrows). At $t=0,T,2T,\cdots$, solid circles (red) indicate the normalized precession axis $\hat{N}$ proportional to $\mathbf{N}$, and they approach to $\mathcal{Z}=0$ monotonically. The horizontal red lines correspond to the threshold $\mathcal{Z} = \mathcal{Z}_\text{th}$. When the trajectory intercepts the red line with  $\ddot{\mathcal{Z}}=\epsilon_x\dot{\mathcal{Y}}>0$ (blue circles), the phase particle is recaptured. By contrast, when the condition $\ddot{\mathcal{Z}}=\epsilon_x\dot{\mathcal{Y}}<0$ is met (dotted circles), $\mathcal{Z}$ keeps on decreasing and flips its sign, allowing the phase particle to roll down. The voltage $V$ jumps exactly when the number of recapturing $n$ decreases [panel (c)].	
	 We use the same parameters in Fig.~\ref{Fig:Fig2dynamics}.
	} \label{Fig:Fig3shapiro}
\end{figure}

{\it DC Shapiro steps---.}
Interestingly, an in-depth analysis of the average voltage $V$ in the regime $I_\text{dc}\geq I_c^*$ reveals a more striking effect: $V$ features sequential sudden steps as the DC current increases. We provide numerical computations of $V$ in Fig.~\ref{Fig:Fig2dynamics}(b) and~\ref{Fig:Fig3shapiro}(a) with realistic experimental parameters~\cite{Bocquillon2017}. The voltage is averaged over a time period $2\times 10^3 \tau_r \sim 160 \, ns$.  In Fig.~\ref{Fig:Fig3shapiro}(a), where we consider a ratio $E_x/E_M =0.67$, more than $4$ sequential jumps are visible. These jumps are the DC Shapiro steps.

To understand their physical origin, we analyze further the mechanism responsible for the rolling of the phase particle down the washboard potential. The key observation is that, for $I_{\rm dc}>I_c^*$, the oscillations of $\mathcal{Z}(t)$ make the washboard potential to keep alternating between configurations (i) and (ii) (described above). As a result it can still develop local minima which temporarily trap the phase particle. The motion of the phase particle thus alternates between oscillations around one local minimum and the rolling from a region $R_j$ down to the next region $R_{j+1}$. The integer number of oscillations within a single region decreases as the average slope of the washboard potential $I_{\rm dc}$ is increased. When this number changes by one, it abruptly modifies the dwelling time $\Omega^{-1}$ of the phase particle within a single region, which causes a sudden jump of the average voltage $V = h\Omega/(2e)$. We observe that, in between the steps, the voltage does not feature flat plateaux, as the frequency $\Omega$ continuously increases with $I_{\rm dc}$, even for a fixed number of oscillations. In the Supplemental Material, we provide short movies which display the time evolution of the system.

Supported by the numerical analysis of the trajectories followed by $\mathbf{R}(t)$ [see for example Fig.~\ref{Fig:Fig3shapiro}(b-d)], we identify the following criteria to understand whether the phase particle is captured at a local minimum or if it rolls down to the next local minimum. For concreteness, we display the passage between regions $R_1$ and $R_2$ in  Fig.~\ref{Fig:Fig2dynamics}(a). When $\mathcal{Z}(t)>\mathcal{Z}_\text{th}$, region $R_1$ features a local minimum, which can trap the phase particle. Once the system dynamics lowers $\mathcal{Z}(t)$ to $\mathcal{Z}(t)=\mathcal{Z}_\text{th}$, the subsequent evolution of the system depends on the sign of the second derivative $\ddot{\mathcal{Z}}=\epsilon_x\dot{\mathcal{Y}}$. If $\ddot{\mathcal{Z}}$ is positive, $\mathcal{Z}(t)$ will rapidly increase back above the threshold and the phase particle will be recaptured by the local minimum in $R_1$. By contrast, if $\ddot{\mathcal{Z}}<0$, $\mathcal{Z}(t)$ will keep on diminishing and eventually flip sign as the phase particle enters the next region $R_2$. The system will then follow an analogous evolution, since Eq.~\eqref{Eq:Washboard} and~\eqref{Eq:MajoranaDynamics} are symmetric under $\mathcal{Y}\to- \mathcal{Y}$,  $\mathcal{Z}\to- \mathcal{Z}$ and $\varphi\to\varphi+2\pi$.

As anticipated before, the steps in the average voltage $V$ are associated with changes in the number $n$ of subsequent recapturing processes as shown in Fig.~\ref{Fig:Fig3shapiro}(b-d). In Fig.~\ref{Fig:Fig3shapiro}(b), the trajectories of $\mathbf{R}(t)$ show that the phase particle is recaptured $n=3$ times before it can roll down to the next region and  $\mathcal{Z}$ can flip sign. In  Fig.~\ref{Fig:Fig3shapiro}(c), where we increase the bias current to exactly match the DC Shapiro step in Fig.~\ref{Fig:Fig3shapiro}(a), we observe a vanishing second derivative $\ddot{\mathcal{Z}}=\epsilon_x\dot{\mathcal{Y}}$ when $\mathcal{Z}$ lowers to $\mathcal{Z}_\text{th}$ for the third time. In Fig.~\ref{Fig:Fig3shapiro}(d), with a higher bias current $I_\text{dc}$, the phase particle is recaptured two times ($n=2$) before rolling down to the next region. With increasing $I_\text{dc}$, two more steps appear when $n=2\rightarrow 1$ and $n=1\rightarrow 0$, and no step appears with higher $I_\text{dc}$.

Finally, we give an estimation of the height of the DC Shapiro steps. To this end, close to the jumps, we roughly approximate the dwelling time within a single region as $\Omega_n^{-1} \sim  n \mathcal{T}$, where $\mathcal{T}$ is the precession period of $\mathbf{R}(t)$. We estimate the latter as $1/\mathcal{T}\sim(E_x^2 E_M^2+E_x^4)^{1/4}/(\pi\hbar)$ by assuming an almost constant $\cos(\varphi) \sim - (I_c^*/I_c)^2$. We can therefore compute the size of the jumps as $\Delta V = (\Omega_{n} -\Omega_{n+1})h/(2e)$, which agrees with the numerical result in Fig.~\ref{Fig:Fig3shapiro}(a). Hence, observations of the size of steps provide a way to estimate the Rabi frequency $1/\mathcal{T}$ of the Majorana qubit by assigning $n$ to the steps and using the formula for $\Delta V$ written above. Notice that $\Delta V$ vanishes when $E_x=0$.

{\it Discussion---.} We have investigated the novel Josephson effect of a TJJ hosting a Majorana qubit. We predict a lowering of the critical current and the emergence of novel DC Shapiro steps. Importantly, the detection of DC Shapiro steps requires experimental techniques similar to the ones used already for the measurements of the known AC Shapiro steps in TJJs~\cite{Wiedenmann2016,Bocquillon2017,Hart2014}.

For clarity, we focus on a specific setup which offers some important features: the localization of the four MBS in different space positions, the absence of other mid-gap states, and the tunability of $E_x$ via the chemical potential $\mu_M$. However, we stress that our findings hold for generic TJJs, provided that they host four MBSs with non-trivial dynamics~\cite{Feng2020}. In Ref.~\cite{SM}, we also have confirmed that our findings are robust against a finite capacitance of the junction or an additional $2\pi$-periodic contribution to the Josephson current.

\begin{acknowledgments}
This work was supported by the W\"urzburg-Dresden Cluster of Excellence on Complexity and Topology in Quantum Matter (EXC2147, project-id 390858490) and by the DFG (SPP1666 and SFB1170 ``ToCoTronics'').
\end{acknowledgments}

\pagebreak
\widetext
\begin{center}
\textbf{\large Supplemental Material  for ``Majorana-induced DC Shapiro steps in topological Josephson junctions"}
\end{center}

This Supplemental Material is organized as follows. In Sec.~I, we derive the inter-edge coupling between Majorana bound states (MBSs) across the finite superconductor. In Sec.~II, we characterize the fixed points reached by the system when the current bias is below the critical value $I_c^*$. The system evolution towards these points is numerically analyzed in Sec.~III. Finally, we show the robustness of the DC Shapiro steps with respect to the presence of a $2\pi$-periodic current and/or a finite capacitance in Sec.~IV.

\section{I. Derivation of the inter-edge coupling}
\begin{figure}[h]
\centering
\includegraphics[width=0.7\columnwidth]{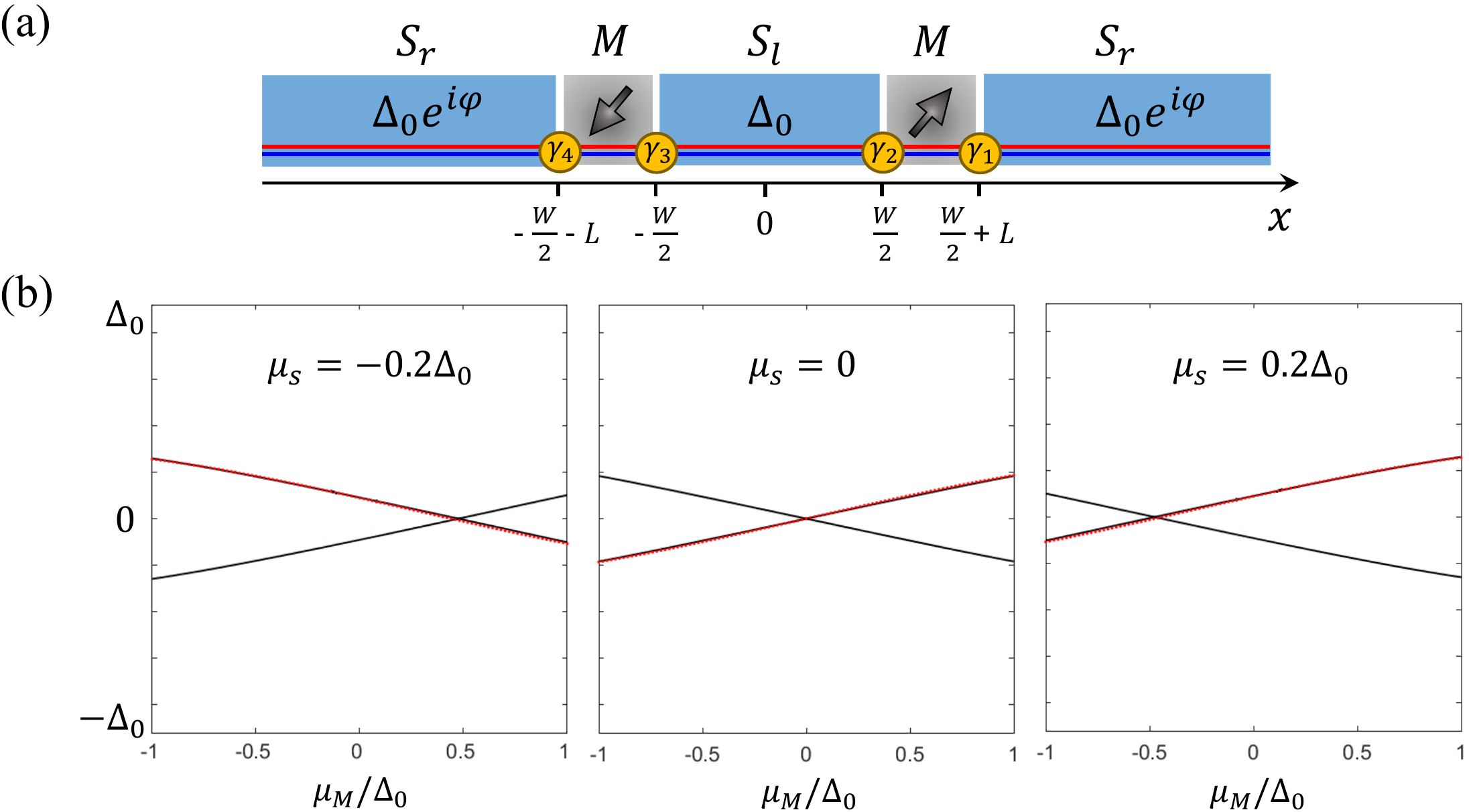}
\caption{(a) Unfolded system, where the coordinate $x$ extends along the upper ($x>0$) and the lower ($x<0$) QSH edge. The length of the QSH edge in the $S_l$-region is $W$, and that in the $M$-region is $L$. The magnetizations at upper and lower edges (gray arrows) are opposite. (b) Comparison of the analytic formula of $E_x$ (red dotted line) and the numerical result of the single particle spectrum (solid line) for various $\mu_S$ with $M=2\Delta_0$. Note that the red dotted line lies exactly on top of one of the two branches of the single particle spectrum. 
} \label{Fig:coord}
\end{figure}

We derive the inter-edge coupling by projecting the full Hamiltonian $H$ into the zero energy wave functions $\zeta_{i}(x)$ of Majorana fermions $\gamma_{i}=\int dx\Psi^\dagger(x)\zeta_i(x)$ for $i=2,3$, i.e., $i E_x=\int dx\zeta_2^\dagger(x)H(x)\zeta_3(x)$. When $\hbar v/L\ll M$ and $\hbar v/W\ll\Delta_0$, we calculate $\zeta_2(x)$ and $\zeta_3(x)$ in the unfolded coordinate system of the U-shaped QSH edge (see Fig.~\ref{Fig:coord}(a)),

\begin{equation}
\zeta_2(x)=
\left\{
\begin{array}{c}

\sqrt{\frac{\Delta_0  \sqrt{M^2-\mu_M ^2}}{2 \hbar v \left(\Delta_0 +\sqrt{M^2-\mu _M^2}\right)}}
e^{-\frac{\sqrt{M^2-\mu_M ^2}}{\hbar v}\left(x-\frac{W}{2}\right)} \left(
\begin{array}{c}
e^{-\frac{i}{2}  (\phi +\chi )} e^{\frac{i \mu_S W}{2\hbar v}}  \\
-i e^{\frac{i}{2} (\phi +\chi )}  e^{-\frac{i \mu_S W}{2\hbar v}}  \\
i e^{-\frac{i}{2} (\phi +\chi )}  e^{\frac{i \mu_S W}{2\hbar v}}  \\
-e^{\frac{i}{2} (\phi +\chi )} e^{-\frac{i \mu_S W}{2\hbar v}}  \\
\end{array}
\right),\,  x>\frac{W}{2} \,(\text{M-region}) \; ,\\

\sqrt{\frac{\Delta_0  \sqrt{M^2-\mu_M ^2}}{2 \hbar v \left(\Delta_0 +\sqrt{M^2-\mu_M ^2}\right)}}
e^{\frac{\Delta_0 }{\hbar v}\left(x-\frac{W}{2}\right)} \left(
\begin{array}{c}
 e^{-\frac{i}{2} (\phi +\chi )} e^{\frac{i \mu_S}{w\hbar v}x} \\
-i e^{\frac{i}{2} (\phi +\chi )} e^{-\frac{i \mu_S}{\hbar v}x}    \\
i e^{-\frac{i}{2} (\phi +\chi )} e^{\frac{i \mu_S}{\hbar v}x}  \\
-e^{\frac{i}{2} (\phi +\chi )} e^{-\frac{i \mu_S}{\hbar v}x}  \\
\end{array}
\right), \qquad x<\frac{W}{2} \,(\text{$S_l$-region})\; ,

\end{array}
\right. \label{Eq:psiSM}
\end{equation}

\begin{equation}
\zeta_3(x)=
\left\{
\begin{array}{c}

\sqrt{\frac{\Delta_0  \sqrt{M^2-\mu_M ^2}}{2 \hbar v \left(\Delta_0 +\sqrt{M^2-\mu_M ^2}\right)}}
e^{-\frac{\Delta_0 }{\hbar v}\left(x+\frac{W}{2}\right)} \left(
\begin{array}{c}
 e^{-\frac{i}{2} (\pi+\phi -\chi )} e^{\frac{i \mu_S}{\hbar v}x}  \\
i e^{\frac{i}{2} (\pi+\phi -\chi )} e^{-\frac{i \mu_S}{\hbar v}x}   \\
-i e^{-\frac{i}{2} (\pi+\phi -\chi )} e^{\frac{i \mu_S}{\hbar v}x}  \\
-e^{\frac{i}{2} (\pi+\phi -\chi )} e^{-\frac{i \mu_S}{\hbar v}x}  \\
\end{array}
\right), \quad x>\frac{-W}{2} \,(\text{$S_l$-region}) \; , \\

\sqrt{\frac{\Delta_0  \sqrt{M^2-\mu_M ^2}}{2 \hbar v \left(\Delta_0 +\sqrt{M^2-\mu_M ^2}\right)}}
e^{\frac{\sqrt{M^2-\mu_M ^2}}{\hbar v}\left(x+\frac{W}{2}\right)} \left(
\begin{array}{c}
e^{-\frac{i}{2}  (\pi+\phi -\chi )} e^{-\frac{i \mu_S W}{2\hbar v}}  \\
i e^{\frac{i}{2} (\pi+\phi -\chi )}  e^{\frac{i \mu_S W}{2\hbar v}} \\
-i e^{-\frac{i}{2} (\pi+\phi -\chi )} e^{-\frac{i \mu_S W}{2\hbar v}} \\
-e^{\frac{i}{2} (\pi+\phi -\chi )} e^{\frac{i \mu_S W}{2\hbar v}} \\
\end{array}
\right),\,  x<-\frac{W}{2} \,(\text{M-region})\; .

\end{array}
\right. \label{Eq:psiMS}
\end{equation}
These wave functions are normalized as $\int_{-\infty}^{\infty}\zeta_i^\dagger(x)\zeta_i(x)dx=1$, $\phi$ is the direction of the magnetization at $x>0$, $\cos\chi=\sqrt{1-(\mu_M/M)^2}$, and $\sin\chi=\mu_M/M$. Notice that the spin direction $\chi$ can be changed by electrically tuning the chemical potential $\mu_M$. We calculate $\int dx\zeta^\dagger_2(x)H(x)\zeta_3(x)$ and obtain $E_x$ in the main text, considering only coupling between adjacent Majorana fermions. We provide the comparison between the analytic formula of $E_x$ and a numerical computation of single particle spectrum in Fig.~\ref{Fig:coord}(b). The numerical result is obtained by putting the QSH system on a lattice model. We implement QSH edge channels on the lattice model which evades the fermion doubling with nonlocal hoppings~\cite{Li2018}.

\section{II. Fixed points of $\varphi_0$ and $\mathcal{Z}_0$ and $I_c^*$}
We first derive $\varphi_0$ and $\mathcal{Z}_0$, and next obtain the analytic expression of $I_c^*$ as the value of $I_\text{dc}$, for which the fixed points become complex numbers.
$\varphi_0$ and $\mathcal{Z}_0$ are derived from the stationary conditions, where time-derivatives vanish,
\begin{equation}
I_\text{dc} = -\mathcal{Z}_0\sin\frac{\varphi_0}{2}, \qquad
\left(
\begin{array}{ccc}
0 & -E_M\cos\frac{\varphi_0}{2} & 0 \\
E_M\cos\frac{\varphi_0}{2} & 0 & -E_x \\
0 & E_x & 0
\end{array}
\right)\left(
\begin{array}{c}
\mathcal{X}_0 \\
\mathcal{Y}_0 \\
\mathcal{Z}_0
\end{array}
\right)=\left(
\begin{array}{c}
0\\
0\\
0
\end{array}
\right).
\end{equation}
As the stationary condition provides four equations for four unknowns of $(\varphi_0, \mathcal{X}_0, \mathcal{Y}_0, \mathcal{Z}_0)$, we can evaluate the fixed points,
\begin{eqnarray}
\cos\varphi_0 &=& -\left(\frac{I_\text{dc}}{I_c}\right)^2 + \sqrt{1+\left(\frac{I_\text{dc}}{I_c}\right)^4 - 2 \left(\frac{I_\text{dc}}{I_c}\right)^2\left(1 + \frac{2E_x^2}{E_M^2}\right)}, \\
\mathcal{Z}_0 &=& -
\sqrt{\frac{1+\left(\frac{I_\text{dc}}{I_c}\right)^2+\sqrt{1+\left(\frac{I_\text{dc}}{I_c}\right)^4-2\left(\frac{I_\text{dc}}{I_c}\right)^2(1+\frac{2E_x^2}{E_M^2})}}{2+\frac{2E_x^2}{E_M^2}}}.
\end{eqnarray}

When the DC current bias $I_\text{dc}$ is lesser than a critical value $I_c^*$, the fixed points of $\varphi_0$ and $\mathcal{Z}_0$ exist as real numbers, otherwise, they are not defined becoming complex numbers. We find that $\varphi_0$ and $\mathcal{Z}_0$ become complex numbers with increasing $I_\text{dc}$ when the argument of the square roots changes sign. Hence, $I_c^*$ is derived from
\begin{equation}
1+\left(\frac{I_c^*}{I_c}\right)^4 - 2 \left(\frac{I_c^*}{I_c}\right)^2\left(1 + \frac{2E_x^2}{E_M^2}\right)=0.
\end{equation}
As $I_c = eE_M/\hbar$ [see the expression for $I_J$ in the main text, before Eq. (3)], we obtain the analytic formula in Eq.~(5) of the main text.

\section{III. Numerical result when $I_\text{dc} \ll I_c$}
\begin{figure}[h]
\centering
\includegraphics[width=0.82\columnwidth]{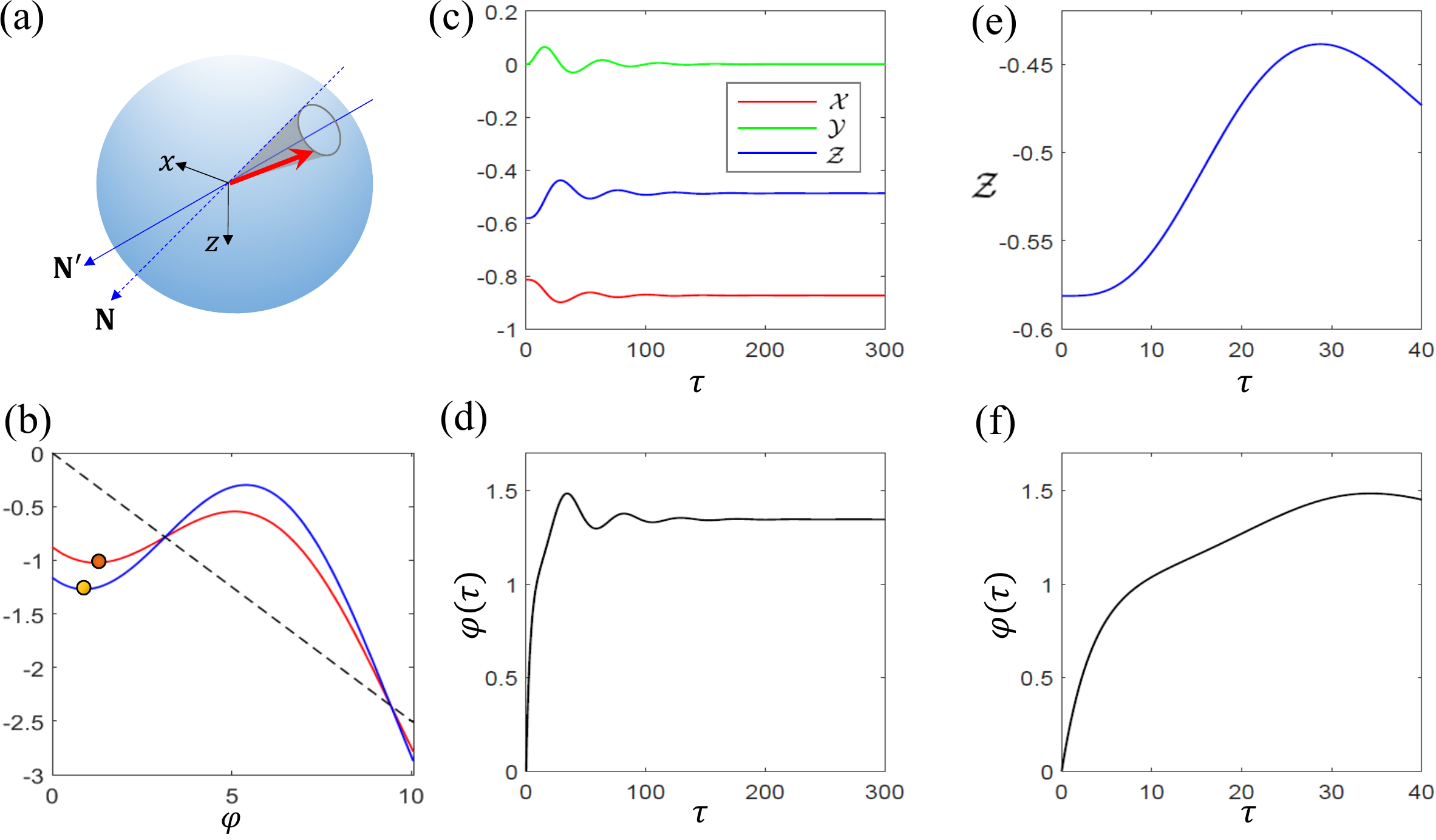}
\caption{The precession of the Majorana qubit (red arrow) on the Bloch sphere (blue shaded circle) is depicted in (a). Before turning on $I_\text{dc}$, the Majorana qubit is aligned along $-\mathbf{N}$, which provides the lowest energy of the junction. After turning on $I_\text{dc}$, $\mathbf{N}$ is tilted to $\mathbf{N}'$, and the Majorana qubit precesses around $\mathbf{N}'$ with the counter-clockwise direction on the surface of the (gray) cone. In (b), dynamics of washboard potential $U(\tau)$ is provided for $i_\text{dc} = 0.3$. Numerical solutions of the Majorana qubit $\mathbf{R}(\tau)$ are presented in (c), while the evolution $\varphi(\tau)$ of the phase particle is shown in (d). Panels in (e) and (f) display behaviors of $\mathcal{Z}$ and $\varphi(\tau)$ around the initial time, respectively. Compared to the initial behavior of $\mathcal{Z}(\tau)$, $\varphi(\tau)$ increases rapidly during the time of the order of $\tau_r$.  We have used $\epsilon_z = 0.05$ and $\epsilon_x = 0.07$.
} \label{Fig:dynamicsSupp}
\end{figure}

We have conducted the numerical calculation of the Josephson effect with  dimensionless parameters.
\begin{eqnarray}
i_\text{dc} &=& -\mathcal{Z}\sin\frac{\varphi}{2} + \frac{d\varphi}{d\tau}, \qquad
\frac{d}{d\tau}
\left(
\begin{array}{c}
\mathcal{X} \\
\mathcal{Y} \\
\mathcal{Z}
\end{array}
\right)
= 2\left(
\begin{array}{ccc}
0 & -\epsilon_z & 0 \\
\epsilon_z & 0 & -\epsilon_x \\
0 & \epsilon_x & 0
\end{array}
\right)
\left(
\begin{array}{c}
\mathcal{X} \\
\mathcal{Y} \\
\mathcal{Z}
\end{array}
\right),
\end{eqnarray}
where $\tau = t/\tau_{r}$, $i_\text{dc}=I_\text{dc}/I_c$, $\epsilon_{x,z}=E_{x,z}\tau_{r}/\hbar$, and $\tau_{r} = \hbar/(2eI_cR_N)$. Notice that $\mathcal{X}$, $\mathcal{Y}$, $\mathcal{Z}$ are dimensionless. When $I_\text{dc} < I_c^*$, $\mathcal{Z}$ does not flip its sign. Then, the dynamics of Majorana qubit and phase particle are quenched as $t\rightarrow\infty$.  We provide the numerical result of the Josephson effect in Fig.~\ref{Fig:dynamicsSupp}. Notice that the Majorana qubit ends up with a superposition of two current-carrying states.

\begin{figure}
	\centering
	\includegraphics[width=.85\linewidth]{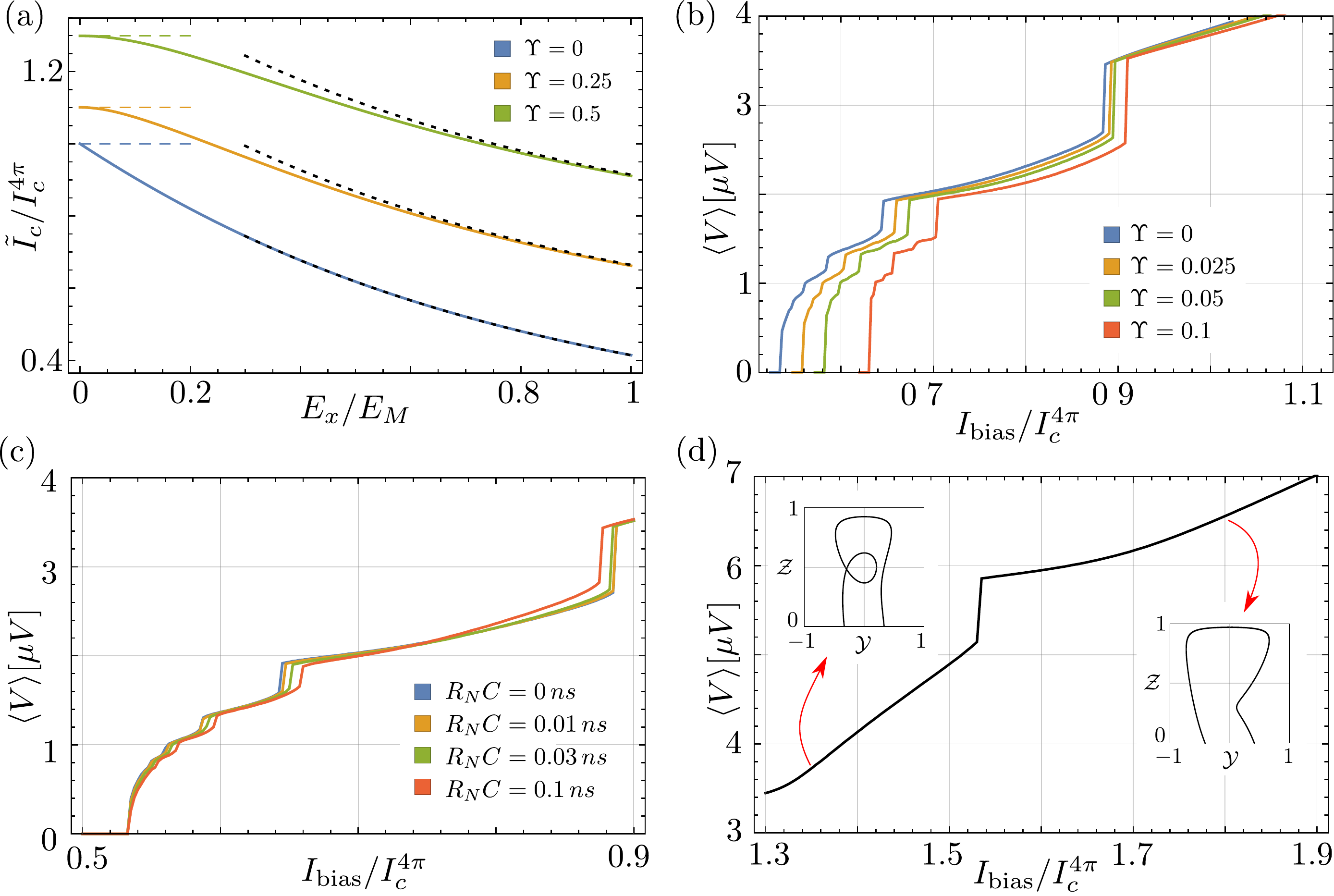}
	\caption{(a) Critical supercurrent $\tilde I_c$ (in units of $I^{4\pi}_c$) as a function of the ratio $E_x/E_M$ for different values of $\Upsilon = I_c^{2\pi}/I^{4\pi}_c$. The horizontal dashed lines show the critical current without inter-edge coupling $\tilde{I}_c^{(0)}$ while the black dotted lines show the approximation $\tilde{I}_c^{(\infty)}$ given in Eq.~\eqref{eq:IcInfty}. (b-c) Time-averaged voltage as a function of the bias current, using the same parameters as in Fig. 3 of the main text, in presence of either finite $2\pi$-periodic supercurrent (b) or finite capacitance (c). (d) Close up of the time-averaged voltage as a function of the bias current, in order to highlight a step. The inset shows two different trajectories of $(\mathcal{Y}(t),\mathcal{Z}(t))$ before and after the step. The parameters in panel (d) are $E_M = 8.3\, \mu V$, $\Upsilon=0.8$, $R_NC = .2 \, ns$ and $2eI^{4\pi}_c R_N/h \sim 2 GHz$.}	
	\label{fig:Ic}
\end{figure}

\section{IV. Effects of capacitance and $2\pi$-periodic supercurrent}
In the main text, we describe the Josephson junction with the RSJ model in Eq.~(3) and show that its interplay with the Majorana dynamics leads to non-trivial phenomena, namely a reduction of the critical current and the appearance of novel DC Shapiro steps. Importantly, we argue that these phenomena persist also when more general terms are added to Eq. (3). In particular, we show that neither the presence of a $2\pi$-periodic supercurrent $I_c^{2\pi}\sin(\phi)$, carried for example by other mid-gap states, nor the existence of a small capacitive coupling $C$ across the junction are detrimental for our findings. The generalization of Eq. (3), which takes into account these additional ingredients, reads
\begin{equation}
I_\text{bias} = -\mathcal{Z}I_c^{4\pi}\sin\frac{\varphi}{2} + \frac{\hbar}{2eR_N}\frac{d\varphi}{dt} + I_c^{2\pi} \sin(\varphi) + C \frac{ \hbar}{2e}\frac{d^2\varphi}{dt^2}.
\end{equation}

The critical supercurrent of the junction, which does not depend on $R_N$ or $C$, is given by the maximal steady current %carried by the ground state
\begin{equation}
\label{eq:tildeIc}
\tilde{I}_c= \max_\varphi \left[
\frac{E_M\cos\frac{\varphi}{2}}{\sqrt{E_x^2 + E_M^2\cos^2\frac{\varphi}{2}}}\; I_c^{4\pi}\sin\frac{\varphi}{2} + I_c^{2\pi} \sin(\varphi),
\right]
\end{equation}
where we have explicitly written the groundstate value of $\mathcal{Z}$ for a given $\varphi$. With a vanishing $I_c^{2\pi}$, one can analytically find
\begin{equation}
\tilde{I}_c^{(I_c^{2\pi}=0)}= I_c^* = I_c^{4\pi} \left(
\sqrt{1+\left(\frac{E_x}{E_M}\right)^2}-\frac{E_x}{E_M}
\right),
\end{equation}
in agreement with the expression for $I_c^*$ given in Eq. (5) of the main text and derived in Section B of the Supplemental Materials. For a finite $2\pi$-periodic current, we provide numerical solutions of Eq.~\eqref{eq:tildeIc} in Fig.~\ref{fig:Ic} (a). Analytical results can be easily derived in the limits $E_x/E_M \to 0$ or $\infty$. In particular, we get
\begin{align}
\label{eq:IcInfty}
\tilde{I}_c^{(0)} &= I_c^{4\pi} \; \frac{3+\sqrt{1+32 \Upsilon^2}}{16 \sqrt{2} \Upsilon }\sqrt{16 \Upsilon^2-1+\sqrt{1+32\Upsilon^2}}\quad \text{with } \Upsilon=\frac{I_c^{2\pi}}{I_c^{4\pi}}\\
\tilde{I}_c^{(\infty)} &= I^*_c + I^{2\pi}_c
\end{align}
which nicely agree with the numerical solutions. We can therefore conclude that the reduction of the critical current induced by the non-trivial Majorana dynamics is robust with respect to the presence of a $2\pi$-periodic supercurrent.

The other panels of Fig. \ref{fig:Ic} show the time-averaged voltage $V$ as a function of the bias current $I_{\rm dc}$, in presence of finite capacitance $C$ and $I^{2\pi}_c$. In particular, in panels (b) and (c) we use the same parameters considered in Fig. 3 of the main text and show that the DC Shapiro steps are robust with respect to either a small finite $\Upsilon$ [panel (b)] or a small finite $C$ [panel (c)]. In panel (d) we consider a different set of parameters to show that, even with a large $\Upsilon=0.8$ and $R_NC = 0.2\; ns$, it is still possible to find finite voltage steps. The small insets show the different trajectories of $(\mathcal{Y}(t),\mathcal{Z}(t))$ before and after the step. While these result do not represent a full characterization of the effects of capacitance and/or $2\pi$-periodic supercurrent, which is beyond the scope of the present work, they clearly show that both the reduction of the critical current and the presence of DC Shapiro steps are robust phenomena.

\end{document}